\documentclass[12pt,epsfig]{article}
\usepackage{amsmath,amssymb,amsthm,color}
\usepackage{graphicx}
\usepackage{cite}
\usepackage{epsfig}
\usepackage{lineno}
\renewcommand{\baselinestretch}{1.75}
\begin{document}
\title{Temperature-induced Coulomb excitations in rhombohedral 3D graphene\\ }
\author{Cheng-Hsueh Yang,$^{a}$ Ting-Wei Jang,$^{b}$ Chang-Ting Liu,$^{c}$ and Chih-Wei Chiu$^{c,*}$\\
\small $^{a}$Department of Physics, National Taiwan University, Taipei, Taiwan\\
\small $^{b}$Department of Civil Engineering, National Cheng Kung University, Tainan, Taiwan\\
\small $^{c}$Department of Physics, National Kaohsiung Normal University, Kaohsiung, Taiwan\\ }
\renewcommand{\baselinestretch}{1}
\maketitle
\renewcommand{\baselinestretch}{1.6}
\begin{abstract}

Low-energy electronic properties of rhombohedral graphite with ABC stacking are studied by the tight-binding model.
There are linear and parabolic bands with and without degeneracy.
They show strongly anisotropic dispersions.
Rhombohedral graphite is a semimetal due to slight overlapping
near the Fermi level between conduction and valence bands.
The interlayer interactions could change the energy dispersion,
state degeneracy, and positions of band-crossing and band-edge state.
Density of states exhibit a shoulder structure, owing to band-edge states near or at high symmetric points.
Low-frequency Coulomb excitation properties with different transferred momenta (${\bf q}'$s) are further studied within the random phase approximation. The Landau dampings is too serious under the parallel transferred momentum (${\bf q}\perp \hat z$); therefore, it is impossible to observe the 3D optical plasmons. However, even for the perpendicular transferred momentum (${\bf q}\parallel \hat z$), the full assistance due to the thermal excitations is necessary to induce the collective charge oscillations along the $z$-axis. The height and position of temperature-induced plasmon peak in the energy loss spectrum are greatly enhanced by the increasing temperature, but weakly depend on the various transferred momenta.
These features are very different from AA- and AB-stacked graphites.

%\vskip 0.6 truecm
%\par\noindent PACS: 78.66.-w, 73.61.Wp, 71.15.Ap
\vskip 0.6 truecm
\par\noindent E-mail address\,:\,giorgio@mail.nknu.edu.tw
%\vskip 0.0 truecm
%\par\noindent $^{b)}$Electronic mails: fl.shyu@msa.hinet.net (F. L. Shyu),  mflin@mail.ncku.edu.tw (M. F. Lin)

\pagebreak
\end{abstract}
\newpage

\section{Introduction}

For three kinds of graphites, there are a lot of theoretical and experimental researches on the electronic properties and Coulomb excitations, clearly illustrating the diversified phenomena by the distinct stacking configurations. According to the first-principles calculations
\cite{PRB46;4540,Carbon32;289,PRB44;13237},
 the AA-, AB- and ABC-stacked graphites, respectively, possess the free carrier densities of
${\sim\,3.5\times\,10^{20}}$ e/cm$^3$, ${\sim\,10^{19}}$ e/cm$^3$, and ${3.0\times\,10^{18}}$ e/cm$^3$, being in sharp contrast with one another.
As a result of the configuration-induced free electrons and holes, such semi-metallic systems are predicted/expected to display the unusual low-frequency single-particle excitations and plasmon modes
\cite{JPSJ69;3781,PRL89;076402,PRB55;13961,PRB69;245419,PRB67;113106,APL89;221910,APL104;161905,JPCM27;125602}.

The simple hexagonal graphite (AA-stacked) exhibits the parallel and the perpendicular collective oscillations relative to the graphitic planes, with frequency higher than 0.5 eV
\cite{PRB44;13237,NJP12;083060,CRCPress;9781138571068}.
Furthermore, certain important differences between two distinct oscillation modes lie in the frequency, intensity and critical momentum, as revealed in Bernal (AB-stacked) graphites
\cite{JPSJ69;3781,APL104;161905,JPCM27;125602}.
Such plasmon modes are strongly modified by the doping effects. The interlayer bondings become weaker in the natural graphite, so that few free carriers only show the lower-frequency plasmons of ${\omega_p\,<0.2}$ eV. In addition to the transferred momentum, the collective excitations are very sensitive to the changes in temperature ($T$)
\cite{PRL66;4,SurSci524;L77}.
They could survive at larger momenta in the increase of temperature, and the frequencies are enhanced by $T$. The $T$-dependent plasmons are the prominent peaks in the energy loss spectra as well as the abrupt edge structures of the optical reflectance spectra.

Up to now, there is absence of theoretical predictions on the low-frequency Coulomb excitations of the rhombohedral (ABC-stacked) graphite. This study will provide the full information from the RPA calculations, e.g., the lowest plasmon frequency among three systems and the most difficult observations using the EELS and optical spectroscopies. After the intercalation of atoms or molecules, the donor-type (acceptor-type) graphite intercalation compounds possess much conduction electrons (valence holes), so that their electrical conductivity might be high as copper
\cite{AdvPhys51;1,PRB81;115428,Carbon57;507,Nanotechnol22;425701}.
The Coulomb excitations of free carriers have been investigated by the 2D superlattice model
\cite{PRB55;13961,PRB34;979},
being responsible for the threshold edge in the measured optical spectra
\cite{PRB138;A197,PRB7;2275,PR178;1340}
and the ${\omega_p\sim\,1}$-eV optical plasmon in the momentum-dependent EELS spectra
\cite{PRL89;076402,PRB138;A197,PRB7;2275,PR178;1340,OptCom1;119,ZPhys243;229,SurSci454;462,PRB38;2112}.

\section{Theories}

The bulk graphites possess the infinite graphene layers, so that their energy bands have an extra wave vector along the $k_z$-direction. Electronic states are described by ${(k_x,k_y,k_z)}$ within the first Brillouin zones [Fig. 1(b)]. Energy dispersions are dominated by the honeycomb lattice on the ${(x,y)}$ plane, stacking configuration; intralayer and interlayer hopping integrals. All the graphites are semi-metals because of the interlayer van der Waals interactions. However, the AA-stacked graphite (ABC-stacked one) exhibits the largest (smallest) overlap between the valence and conduction bands and thus the highest (lowest) free electron and hole densities, directly reflecting the geometric symmetry. Band structures and free carrier densities are quite different for three kinds of graphites, and electronic excitations are expected to behave so
\cite{JPSJ69;3781,JPSJ70;897,JPSJ81;104703}.
For example, the low-frequency plasmon due to the interlayer atomic interactions are very sensitive to AA, AB or ABC stacking.

It is relatively easy to utilize the conventional primitive unit cell to fully explore the essential physical properties, while the opposite for the primitive rhombohedral unit cell except for few cases \cite{CRCPress;9781138571068}.
As a result, the Bloch wave functions are composed of the six tight-binding functions on three layers. Similar to the calculations of trilayer graphene systems, the ${6\times\,6}$ Hamiltonian matrix could be reduced to consist of two independent block matrices:

\begin{equation}
H=\left(
\begin{array}{ccc}
H_{1} & H_{2} & H_{2}^{\star}\\
H_{2}^{\star} & H_{1} & H_{2}\\
H_{2} & H_{2}^{\star} & H_{1}\\
\end{array}%
\right)
\end{equation}%
where ${2\times\,2}$ $H_{1}$ and $H_{2}$ matrix take the analytic forms in their elements forms
\begin{equation}
H_{1}=\left(
\begin{array}{cc}
0 & \beta_{0}h(k_{x},k_{y})\\
\beta_{0}h^{*}(k_{x},k_{y}) & 0\\
\end{array}%
\right)
\end{equation}%
and
\begin{eqnarray}
H_{2}=\,\,\,\,\,\,\,\,\,\,\,\,\,\,\,\,\,\,\,\,\,\,\,\,\,\,\,\,\,\,\,\,\,\,\,\,\,\,\,\,\,\,\,
\,\,\,\,\,\,\,\,\,\,\,\,\,\,\,\,\,\,\,\,\,\,\,\,\,\,\,\,\,\,\,\,\,\,\,\,\,\,\,\,\,\,\,
\,\,\,\,\,\,\,\,\,\,\,\,\,\,\,\,\,\,\,\,\,\,\,\,\,\,\,\,\,\,\,\,\,\,\,\,\,\,\,\,\,\,\,
\,\,\,\,\,\,\,\,\,\,\,\,\,\,\,\,\,\,\,\,\,\,\,\,\,\,\,\,\,\,\,\,\,\,\,\,\,\,\,\,\,\,\,
\,\,\,\,\,\,\,\,\,\,\,\,\,\,\,\,\,\,\,\,\,\,\,\cr
\left(
\begin{array}{cc}
(\beta_{4}\exp(ik_{z}I_{z})+\beta_{5}\exp(-i2k_{z}I_{z}))h^{*}(k_{x},k_{y})
& \beta_{1}\exp(ik_{z}I_{z}) +\beta_{2}\exp(i2k_{z}I_{z}) \\
(\beta_{3}\exp(ik_{z}I_{z})+\beta_{5}\exp(-i2k_{z}I_{z}))h(k_{x},k_{y})
&(\beta_{4}\exp(ik_{z}I_{z})+\beta_{5}\exp(-i2k_{z}I_{z}))h^{*}(k_{x},k_{y})\\
\end{array}%
\right)
\end{eqnarray}%
All the significant hopping integrals between two different sublattices on three layers, as clearly indicated in Fig. 1(a), cover $\beta_0$=3.16 eV, $\beta_1$=0.36 eV, $\beta_2$=$-$ 0.02 eV,
$\beta_3$=0.32 eV, $\beta_4$=$-$0.03 eV, and $\beta_5$=0.013 eV
\cite{PRB46;4540},
in which their definitions are similar to those in the AB-stacked graphite. This method for the band-structure calculations is also successful to use in the other systems with different dimensions \cite{NJP12;083060,OE19;23350,JAP103;073709,PCCP18;7573,PLA352;446,PhysicaE40;1722,PRB70;075411}.

The 3D transferred momentum (${q_x,q_y,q_z}$) in graphites is conserved during the electron-electron Coulomb intercations, as observed in 3D electron gas. The analytic form of the dielectric function, which is similar for any graphites, is directly evaluated from the RPA

\begin{eqnarray}
\epsilon(q_{x},q_{y},q_{z},\omega\,)=\epsilon_{0}-\sum_{h,h^{\prime}=c,v} \int_{1st BZ}\frac{e^{2}d^{2}{\mathbf k}_{\parallel}dk_{z}}{q^{2}\pi^{2}}|\langle {\mathbf k}_{\parallel}+{\mathbf q}_{\parallel},k_{z}+q_{z};h^{\prime}|e^{i{{\mathbf q}\cdot {\mathbf r}}}|{\mathbf k}_{\parallel},k_{z};h\rangle|^{2}\cr
\times \frac{f(E^{h^{\prime}}({\mathbf k}_{\parallel}+{\mathbf q}_{\parallel},k_{z}+q_{z}))-
f(E^{h}({\mathbf k}_{\parallel},k_{z}))}
{E^{h^{\prime}}({\mathbf k}_{\parallel}+{\mathbf q}_{\parallel},k_{z}+q_{z})-
E^{h}({\mathbf k}_{\parallel},k_{z})-(\omega+i\Gamma)}.
\end{eqnarray}
The $k_z$-integration of the first Brillouin zone is distinct in three kinds of stackings configurations. There exist the low-frequency and $\pi$ plasmons. The former could survive at small transfer momenta, while it might be difficult to observe it at large ones. Under this case, the anisotropic dependence on the ${(q_x,q_y)}-$ plane is negligible; that is, ${{\bf q}\,=[{\bf q_{\parallel}}\,,q_z]}$. Since the low-lying energy bands are almost isotropic near the K/K$^\prime$ point, the 3D integration could be reduced to the 2D integration, i.e., ${\int_{1stBZ}\,d^2{\bf k}dk_z}$ $\to$${\int_{1stBZ}\,2\pi\,k_{\parallel}dk_{\parallel}dk_z}$.

\section{Results}

Using the hexagonal unit cell, the rhombohedral graphite with ABC stacking , as show in Fig. 1(c) and 1(d), has three pairs of valence and conduction bands, mainly owing to the
zone-folding effect on the $\pi$-electronic structure. For the low-lying electronic structure, the ${k_z}$-dependent energy width is less than ${10}$meV [Fig. 1(c)], indicating the
strong competitions among the interlayer atomic intercations in the specific ABC stacking. Band overlap is almost vanishing, so that the 3D density of free electrons and holes is
lowest (highest) for the ABC stacking (the AAA stacking). There is one non-degenerate Dirac-cone structure and two degenerate parabolic bands near the K and H points [Fig. 1(d)].
The similar state degeneracy appears at the middle energies close to the M and L points, in which the energy widths of the interlayer-interaction-induced saddle points are directly
reflected in DOS [Fig. 2(c)]. In fact, only one pair of ${2p_z}$-dominated valence and conduction bands is revealed under a rhombohedral unit cell
\cite{NJP15;053032}.
A three-dimensional Dirac-cone structure is composed of the tilted anisotropic Dirac cones around the spirally located Dirac points. Moreover, the Dirac points form a nodal spiral
in wave vector space due to the accidental degeneracy, which can only be realized in rhombohedral graphite up to now.

The main features of DOS in the ABC-stacked graphite are unique [Fig. 2(c)], being thoroughly different from those in the AA- and AB-stacked graphites [Figs. 2(a) and 2(b)]. A
almost symmetric V-shape structure [the inset in Fig. 2(c)], which purely comes from the 3D Dirac-cone structure, is initiated from the Fermi level
\cite{CRCPress;9781138571068}.
The DOS at $E_F$ is smallest among three kinds of graphites, further illustrating the weakest interlayer atomic interactions after the strong competition. The low-frequency
collective excitations are expected to be observable only under the sufficiently high temperature.
As the state energies gradually grow, a pair of finite-width shoulder structures appear at 0.11 $\beta_0$ and ${-0.12}$ $\beta_0$. That the anisotropic linear energy dispersions are
getting into the parabolic forms is the main reason. The critical points are, respectively,  [Dirac points $\&$ extreme points] and saddle points at low and middle frequencies. The
latter lead to the very sharp peaks at 0.959 $\beta_0$, 1.010 $\beta_0$ and $-$0.991 $\beta_0$, in which they are similar to the 2D logarithmic-form symmetric structures. The
conduction and valence DOSs present the slightly splitting peaks and a single peak, respectively. The strong symmetric peaks and the V-shape form, which are, respectively, located
at low and middle frequencies, clearly indicate the quasi-2D behaviors
\cite{CRCPress;9781138571068}.

The lower-symmetry rhombohedral graphite has the lowest the free electron/hole density among three kinds of well-behaved stacking configurations, so that it is expected to be
relatively difficult in creating the low-frequency collective excitations.
Most of its excitation phenomena are similar those of the Bernal graphite
\cite{JPSJ69;3781}.
For example, the Landau dampings is too serious under the parallel transferred momentum; therefore, it is impossible to observe the 3D optical plasmons. However, even for the perpendicular transferred momentum, the full assistance due to the thermal excitations is necessary to induce the collective charge oscillations along the $z$-axis [Figs.3 \& 4].
At ${T=600}$ K, the maximum value of Im${[\epsilon\,]}$/Re${[\epsilon\,]}$ drops from ${\sim\,2000}$ to ${\sim\,200}$ as $q_z$ grows from 0.001 ${\AA^{-1}}$ to 0.03 ${\AA^{-1}}$
[Figs. 3(a) and 3(b)].
The drastic change in the dielectric function hardly affects the height of plasmon peak [Fig. 3(c)], since the latter is determined by the second zero point of Re${[\epsilon\,]}$
and the value of Im${[\epsilon\,]}$. The peak positions only weakly depend on the various transferred momenta, directly reflecting the very weak energy dispersion along the KH line
[Fig. 1(d)].
Moreover, the bare and screened response functions are very sensitive to the change in temperature. Apparently, the maximum strength of the e-h excitations [Fig. 3(d)], the
frequency corresponding to the second zero point in Re${[\epsilon\,]}$ [Fig. 4(a)], and the height of plasmon peak in the energy loss spectrum [Fig. 4(b)] are greatly enhanced by
the increasing temperature.

The dispersion relations between the free-carrier-induced plasmon frequency and the transferred momentum/the temperature could provide the first-step information about the
experimental examinations. The plasmon frequency lies in the range of ${0.02 \beta_0\,<\omega_p\,<0.04 \beta_0}$ during the variation of momentum or temperature, as clearly shown in
Figs. 5(a) and 5(b).
In general, the $q$-dependence is weak, directly reflecting a very narrow band width along the KH line [Figs. 1(c) and 1(d)]. The low-frequency optical plasmon modes could survive
only at finite temperature even under ${q_z\rightarrow\,0}$, being thoroughly different from those in AB-stacked graphite
\cite{JPSJ69;3781}.
Furthermore, the critical momentum grows with the increasing temperature [Fig. 5(a)]. The plasmon frequency strongly depends on temperature at very small $q$$^{,}$s, while the
$T$-dependence becomes negligible for other cases. The temperature-created plasmons in rhombohedral graphite could be verified by the high-resolution energy loss spectra
\cite{PRL89;076402,PRB138;A197,PRB7;2275,PR178;1340,OptCom1;119,ZPhys243;229,SurSci454;462,PRB38;2112}
and the optical measurements, as done for Bernal graphite
\cite{PRB138;A197,PRB7;2275,PR178;1340,NanoRes10;234}.

\section{Conclusions}

The tight-binding model and the random phase approximation are respectively utilized to study the low-frequency electronic and Coulomb-excitation properties of ABC-stacked graphite.
In conclusion, it can be said that the strength (the strength and the frequency) of the single-particle excitations strongly depends on the temperature $T$ (the magnitude of the perpendicular transferred momentum $q_z$).
Because of the energy dispersion along the KH direction and the Fermi–Dirac distribution, the strength and the frequency declines and grows with the increase of $q_z$, respectively. Moreover, due to the increase of $T$, the raised free electrons and holes enhance the strength of the single-particle excitations. As to the collective excitations, the optical plasmons under the perpendicular transferred momentum are very sensitive to temperature. The plasmon modes at long wavelength limit are clearly revealed in the energy loss spectra, only when the thermal excitations are required to supply the sufficient free electrons and holes, e.g., the existence of collective excitations at ${\bf q\rightarrow\,0}$ only for ${T>100}$ K. In general, the higher the temperature (the transferred momentum) is, the larger (the higher) the critical momentum (temperature) is. Up to now, there are no experimental examinations on the screened excitation spectra of the ABC-stacked graphites.

\begin{figure}[p]
%h=here, t=top, b=bottom, p=separate figure page
\begin{center}\leavevmode
\rotatebox{0}{\includegraphics[width=1.0\linewidth]{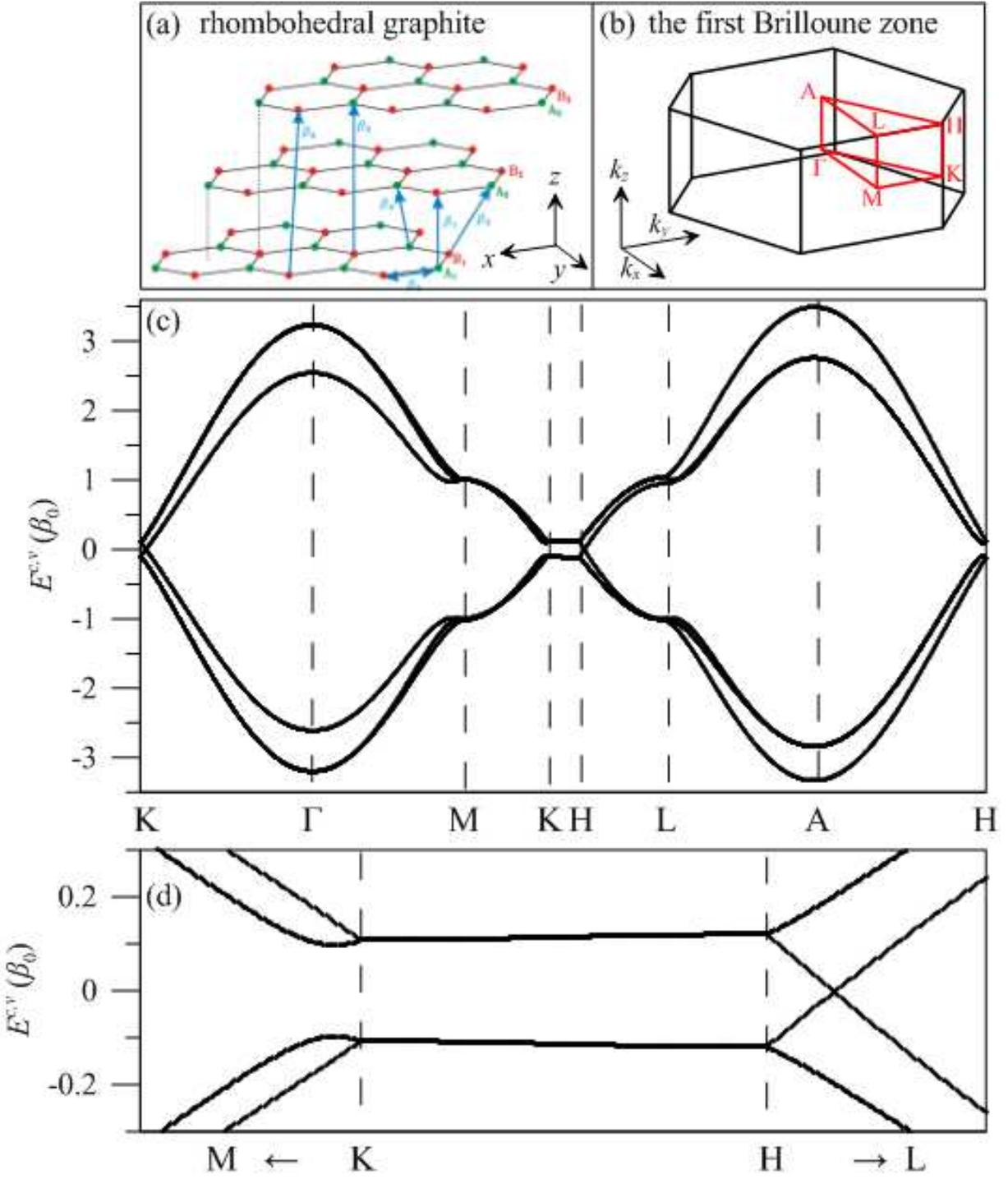}}
%{\includegraphics[width=2\linewidth]{fig1.eps}}
\caption{The geometric structure (a), the first Brilloune zone (b), and the whole-energy/low-lying band structures  (c)/(d) of the rhombohedral graphite.}
\label{}\end{center}\end{figure}

\begin{figure}[p]
%h=here, t=top, b=bottom, p=separate figure page
\begin{center}\leavevmode
\rotatebox{0}{\includegraphics[width=0.95\linewidth]{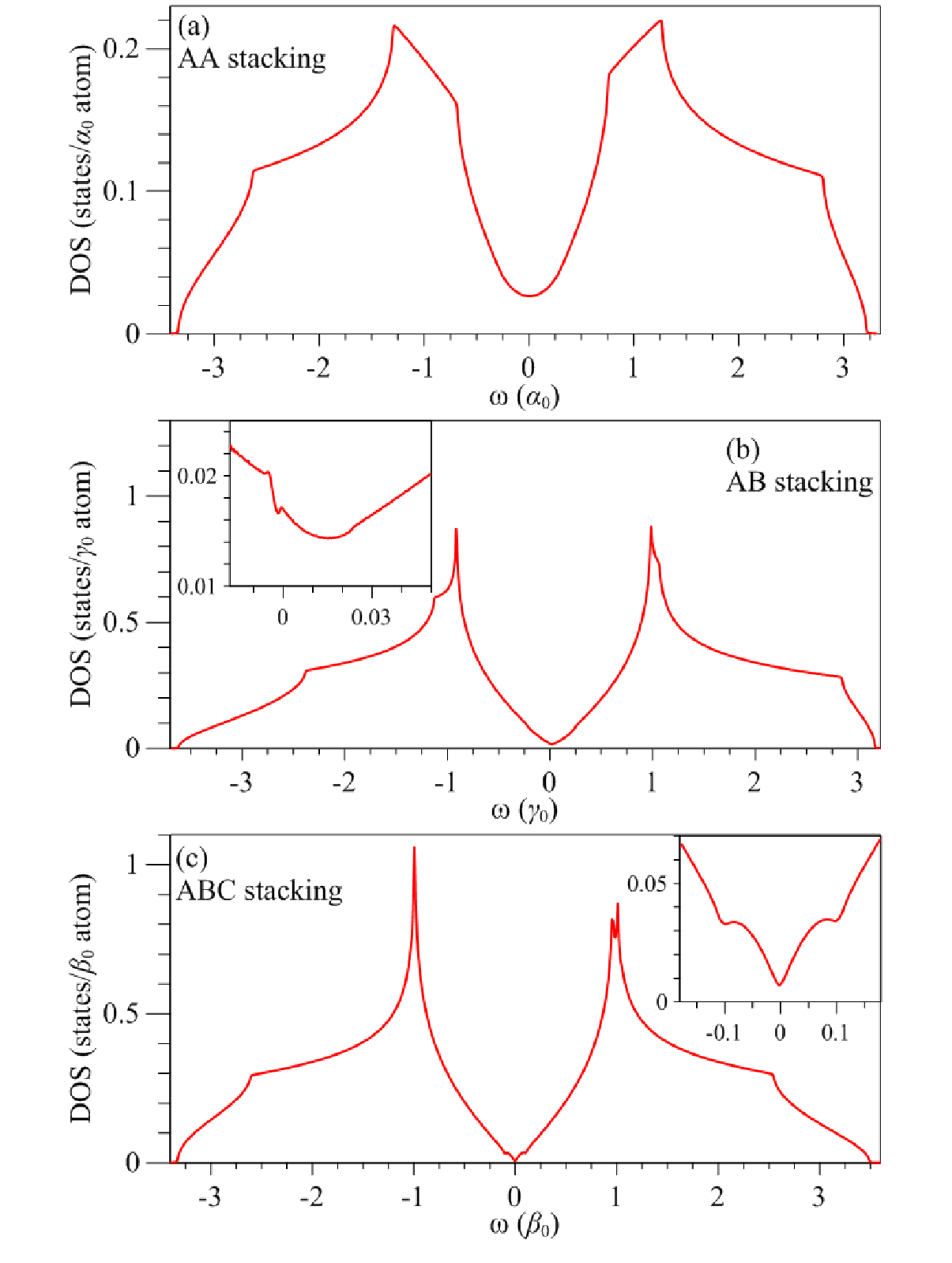}}
%{\includegraphics[width=2\linewidth]{fig1.eps}}
\caption{The density of states for the (a) AA- (b) AB- and (c) ABC-stacked graphites. The insets show the van Hove singularities near the Fermi level.}
\label{}\end{center}\end{figure}

\begin{figure}[p]
%h=here, t=top, b=bottom, p=separate figure page
\begin{center}\leavevmode
\rotatebox{0}{\includegraphics[width=14.5 cm]{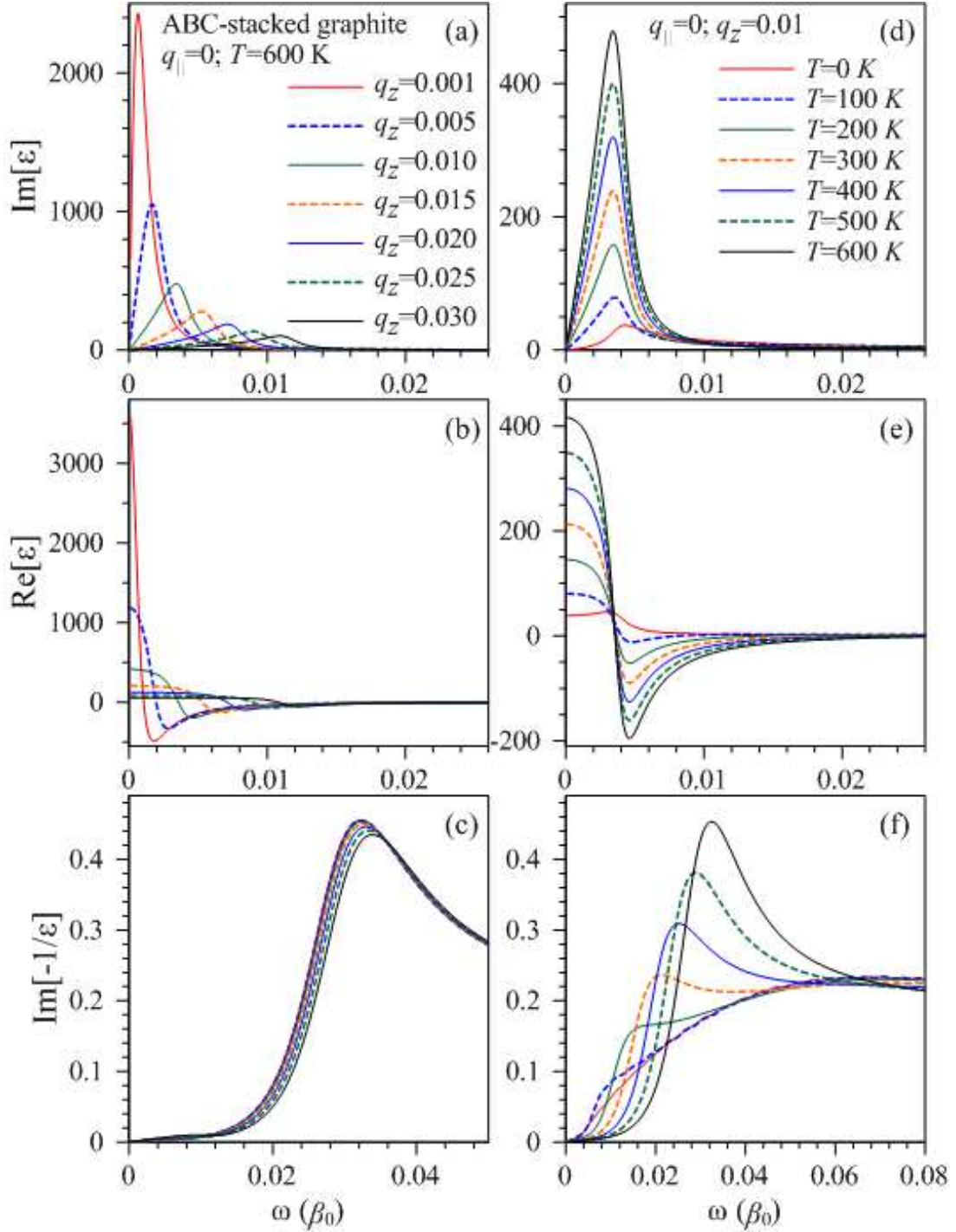}}
%{\includegraphics[width=2\linewidth]{fig1.eps}}
\caption{The imaginary and real part of the low-frequency dielectric function for the ABC-stacked graphite, being shown in [(a), (b)]/[(c), (d)], respectively, under the conditions: $T$=600 K $\&$ various $q_z$s, and ${q_z}$=0.01 ${\AA^{-1}}$ and different temperatures.}
\label{}\end{center}\end{figure}

\begin{figure}[p]
%h=here, t=top, b=bottom, p=separate figure page
\begin{center}\leavevmode
\rotatebox{0}{\includegraphics[width=14.5 cm]{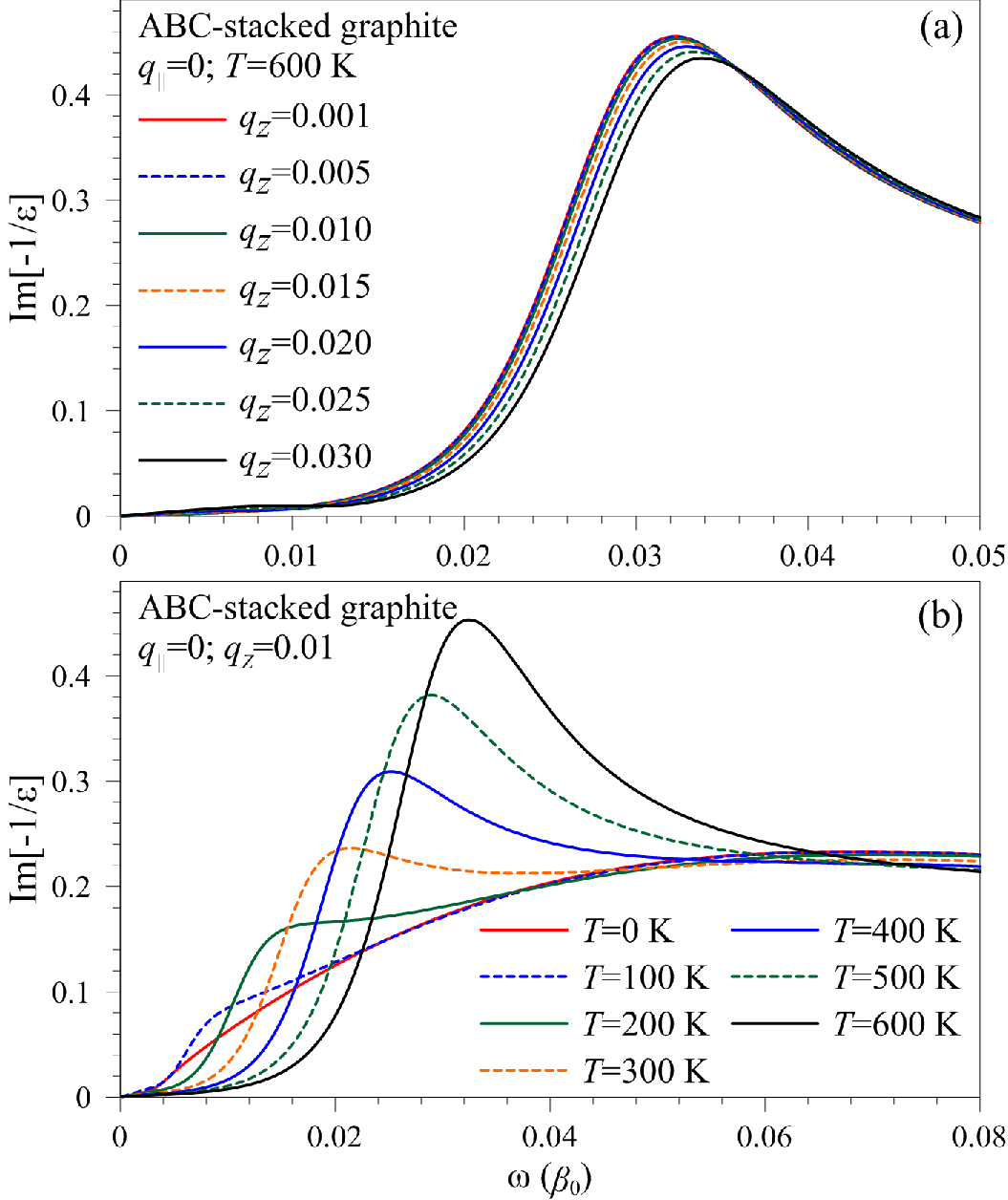}}
%{\includegraphics[width=2\linewidth]{fig1.eps}}
\caption{The energy loss function for the ABC-stacked graphite under the conditions: $T$=600 K $\&$ various $q_z$s, and ${q_z}$=0.01 ${\AA^{-1}}$ and different temperatures, being shown in (a) and (b), respectively.}
\label{}\end{center}\end{figure}

\begin{figure}[p]
%h=here, t=top, b=bottom, p=separate figure page
\begin{center}\leavevmode
\rotatebox{0}{\includegraphics[width=1.0\linewidth]{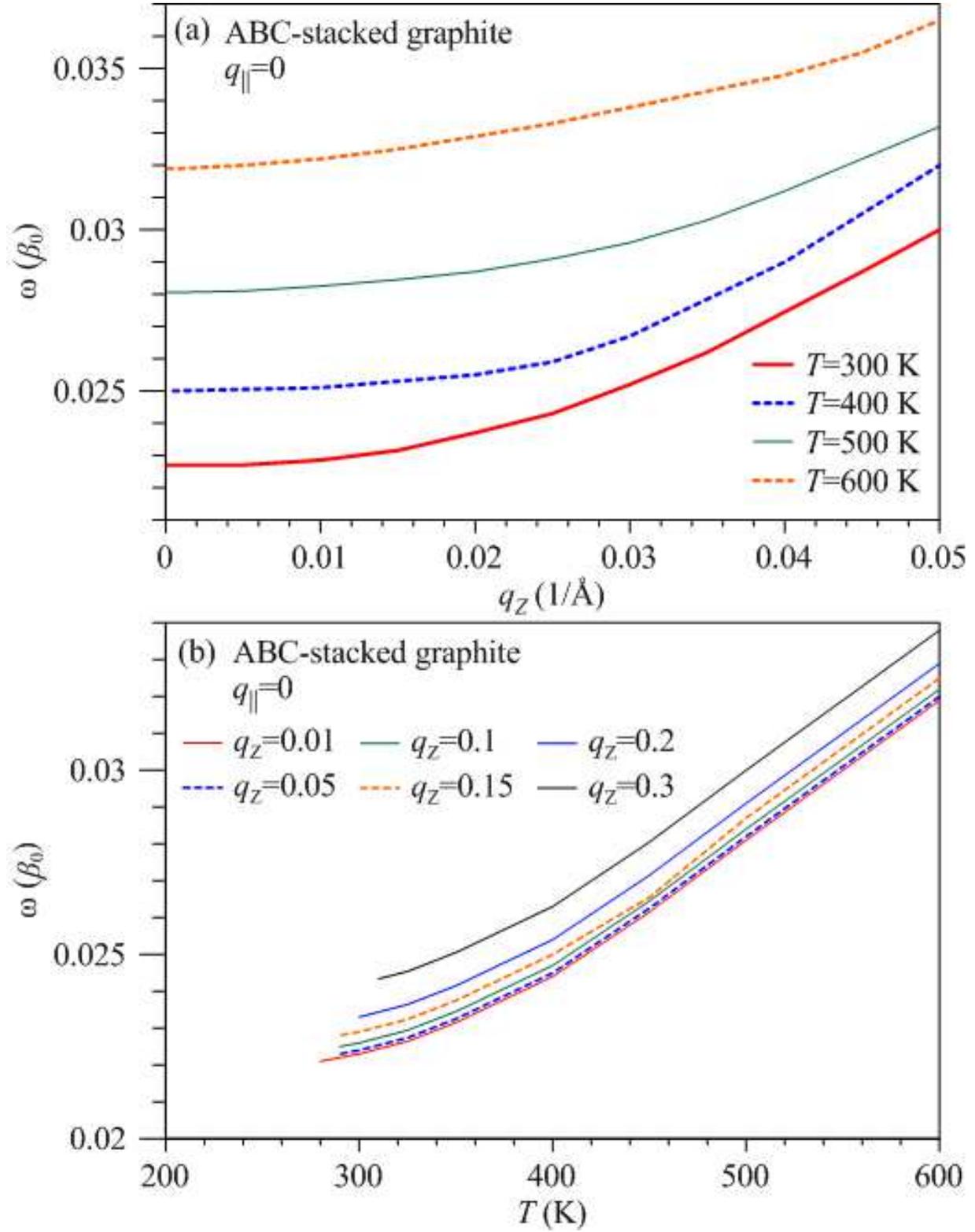}}
%{\includegraphics[width=2\linewidth]{fig1.eps}}
\caption{The dispersion relations of the plasmon frequencies on (a) the transferred momentum and
(b) the temperature, respectively, under the specific $T$$^{,}$s and $q_z$$^{,}$s.}
\label{}\end{center}\end{figure}

\centerline {\bf Acknowledge}

We would like to acknowledge the financial support from the Ministry of Science and Technology of the Republic of China (Taiwan) under Grant No. MOST 107-2112-M-017-001.

\end{document}